\documentclass[preprintnumbers,prd,twocolumn,showpacs,floatfix,preprintnumbers,letterpaper,nofootinbib,amsmath,amssymb,superscriptaddress]{revtex4}
\usepackage{graphicx}
\usepackage{epsfig}
\usepackage{bm}
\usepackage{amsfonts}

\usepackage{color}

\newcommand{\rhophi}{\rho_{\rm DDE}}

\newbox\pippobox

\def\bx{{\bf x}}
\def\bk{{\bf k}}

\def\be{\begin{equation}}
\def\ee{\end{equation}}
\def\ba{\begin{eqnarray}}
\def\ea{\end{eqnarray}}

\newcommand {\lla} {\ {\raise-.5ex\hbox{$\buildrel<\over\sim$}}\ }
\renewcommand{\(}{\left(}
\renewcommand{\)}{\right)}
\renewcommand{\[}{\left[}
\renewcommand{\]}{\right]}

\usepackage[T1]{fontenc}
\usepackage[latin1]{inputenc}
\usepackage{graphicx}
\usepackage[english]{babel}
\usepackage{amsmath}
\usepackage{amssymb}
\usepackage{amsfonts}

\begin{document}

\title{$f(T)$ gravity mimicking dynamical dark energy. Background and perturbation analysis}

\author{James~B.~Dent} \email{jbdent@asu.edu}
\affiliation{Department of Physics and School of Earth and Space
Exploration, Arizona State University, Tempe, AZ 85287-1404}

\author{Sourish Dutta}
\email{sourish.d@gmail.com} \affiliation{Department of Physics and
Astronomy, Vanderbilt University, Nashville, TN  ~~37235}

\author{Emmanuel N. Saridakis}
\email{msaridak@phys.uoa.gr}
 \affiliation{College of Mathematics
and Physics,\\ Chongqing University of Posts and
Telecommunications, Chongqing 400065, P.R. China }

\begin{abstract}
We investigate $f(T)$ cosmology in both the background, as well as
in the perturbation level, and we present the general formalism
for reconstructing the equivalent one-parameter family of $f(T)$
models for any given dynamical dark energy scenario. Despite the
completely indistinguishable background behavior, the
perturbations break this degeneracy and the growth histories of
all these models differ from one another. As an application we
reconstruct the $f(T)$ equivalent for quintessence, and we show
that the deviation of the matter overdensity evolution is strong
for small scales and weak for large scales, while it is negligible
for large redshifts.
\end{abstract}


\maketitle

\section{Introduction}

Cosmological data from a wide range of sources including type Ia
supernovae \cite{union08, perivol, hicken}, the cosmic microwave
background \cite{Komatsu}, baryon acoustic oscillations
\cite{bao,percival}, cluster gas fractions
\cite{Samushia2007,Ettori} and gamma ray bursts
\cite{Wang,Samushia2009} seem to indicate that at least 70\% of
the energy density in the universe is in the form of an exotic,
negative-pressure component, which drives the universe
acceleration. Although the simplest way to explain this behavior
is the consideration of a cosmological constant \cite{c7}, the
known fine-tuning problem led to the dark energy paradigm. The
dynamical nature of dark energy, at least in an effective level,
can originate from a variable cosmological ``constant''
\cite{varcc}, or from various fields, such is a canonical scalar
field (quintessence)
\cite{RatraPeebles,ZlatevWangSteinhardt,SteinhardtWangZlatev:1999b,WangSteinhardt:1998,quint},
a phantom field, that is a scalar field with a negative sign of
the kinetic term \cite{phant}, the combination of quintessence and
phantom in a unified model named quintom \cite{quintom}, or from
k-essence \cite{coy} and unparticles \cite{unparticleDE}. On the
other hand, the dynamical dark energy can effectively be
described by modifying gravity itself, using a functions of the
curvature scalar \cite{Carroll:2003wy},  of the Gauss-Bonnet
invariant \cite{GB}, or of the square of the Weyl tensor
\cite{conformalgravDE},
 higher derivatives in the
action \cite{Nojiri:2005jg}, braneworld extensions \cite{brane},
string-inspired constructions \cite{string}, holographic
properties \cite{holoext}, UV modifications \cite{Horawa} etc.

An interesting alternative to General Relativity is the so-called
$f(T)$ gravity, which has recently received considerable attention
as a possible explanation of the late-time acceleration
\cite{Ferraro:2006jd,Bengochea:2008gz,Linder:2010py}. It is based
on the old idea of the  ``teleparallel'' equivalence of General
Relativity (TEGR) \cite{ein28,Hayashi79}, which, instead of using
the Reimann-Cartan space with curvature but no torsion defined via
the Levi-Civita connection, uses the Weitzenb{\"o}ck connection
that has no curvature but only torsion. The dynamical objects in
such a framework are the four linearly independent vierbeins. The
advantage of this framework is that the torsion tensor is formed
solely from products of first derivatives of the tetrad.  As
described in \cite{Hayashi79}, the Lagrangian density, $T$, can
then be constructed from this torsion tensor under the assumptions
of invariance under general coordinate transformations, global
Lorentz transformations, and the parity operation, along with
requiring the Lagrangian density to be second order in the torsion
tensor. However, instead of using the tensor scalar $T$ the
authors of \cite{Bengochea:2008gz,Linder:2010py} generalized the
above formalism to a modified $f(T)$ version, thus making the
Lagrangian density a function of $T$, similar to the well-known
extension of $f(R)$ Einstein-Hilbert action. In comparison with
$f(R)$ gravity, whose fourth-order equations may lead to
pathologies, $f(T)$ gravity has the significant advantage of
producing field equations which are at most second order in field
derivatives. This feature has led to a rapidly increasing interest
in the literature, and apart from obtaining acceleration
\cite{Bengochea:2008gz,Linder:2010py} one can reconstruct a
variety of cosmological evolutions
\cite{Myrzakulov:2010vz,Myrzakulov:2010tc}, add a scalar field
\cite{Yerzhanov:2010vu}, examine the conformal transformations
\cite{Yang:2010ji}, use observational data in order to constrain
the model parameters \cite{Wu:2010xk}, examine the dynamical
behavior of the scenario \cite{Wu:2010mn} and the possibility of
the phantom divide crossing \cite{Wu:2010av,Bamba:2010iw}, and
proceed beyond the background evolution, investigating the vacuum
and matter perturbations \cite{Dent:2010va}.

In this paper we are interested in constructing $f(T)$ scenarios,
that exhibit the same background behavior with any given
cosmological model. However, we additionally investigate the
perturbation evolution, and in particular we examine the growth of
matter overdensity, since it distinguishes between the dynamical
dark energy scenario and its equivalent family of $f(T)$ models.

The layout of this paper is as follows. In Section \ref{model} we
briefly review the cosmology of $f(T)$ gravity,  both at the
background and linearized regimes. In Section \ref{Reconstr} we
present the general formalism of reconstructing the equivalent
$f(T)$ models of any given dynamical dark energy scenario, as well
as the formalism of using perturbations in order to distinguish
them. In Section \ref{Quintappl} we apply the obtained formalism
in the quintessence scenario. Finally, Section \ref{conclusions}
summarizes our results.

\section{$f(T)$ gravity and cosmology}
\label{model}

In this section we present $f(T)$ gravity, we provide the
background cosmological equations in a universe governed by such a
modified gravitational sector, and we give the first order
perturbed equations. Throughout the work we consider a flat
Friedmann-Robertson-Walker (FRW) background geometry with metric
\begin{equation}
ds^2= dt^2-a^2(t)\,\delta_{ij} dx^i dx^j,
\end{equation}
where $a(t)$ is the scale factor. In this manuscript our notation
is as follows: Greek indices $\mu, \nu,$... run over all
coordinate space-time 0, 1, 2, 3, lower case Latin indices (from
the middle of the alphabet) $i, j, ...$  run over spatial
coordinates 1, 2, 3, capital Latin indices $A, B, $... run over
the tangent space-time 0, 1, 2, 3, and lower case Latin indices
(from the beginning of the alphabet) $a,b, $... will run over the
tangent space spatial coordinates 1, 2, 3.

\subsection{$f(T)$ gravity}
\label{fTgrav}

Let us present $f(T)$ gravity. As stated in the Introduction, the
dynamical variable of the old ``teleparallel'' gravity, as well as
its $f(T)$ extension, is the vierbein field
${\mathbf{e}_A(x^\mu)}$, where capital as well as Greek indices
take the values $0, 1, 2, 3$. This forms an orthonormal basis for
the tangent space at each point $x^\mu$ of the manifold, that is
$\mathbf{e} _A\cdot\mathbf{e}_B=\eta_{AB}$, where $\eta_{AB}=diag
(1,-1,-1,-1)$. Furthermore, the
  vector $\mathbf{e}_A$ can be analyzed with the use of its components $e_A^\mu$
 in a coordinate basis, that is
$\mathbf{e}_A=e^\mu_A\partial_\mu $.

In such a framework, the metric tensor is obtained from the dual
vierbein as
\begin{equation}
\label{metrdef} g_{\mu\nu}(x)=\eta_{AB}\, e^A_\mu (x)\, e^B_\nu
(x).
\end{equation}
Contrary to General Relativity, which uses the torsionless
Levi-Civita connection, in the present formalism ones uses the
curvatureless Weitzenb\"{o}ck connection \cite{Weitzenb23}, whose
 torsion tensor reads
\begin{equation}
 \label{torsion2}
{T}^\lambda_{\:\mu\nu}=\overset{\mathbf{w}}{\Gamma}^\lambda_{\nu\mu}-\overset
{\mathbf{w}}{\Gamma}^\lambda_{\mu\nu}=e^\lambda_A\:(\partial_\mu
e^A_\nu-\partial_\nu e^A_\mu).
\end{equation}
Moreover, the contorsion tensor, which equals the difference
between Weitzenb\"{o}ck and Levi-Civita connections, is defined as
\begin{equation}
 \label{cotorsion}
K^{\mu\nu}_{\:\:\:\:\rho}=-\frac{1}{2}\Big(T^{\mu\nu}_{\:\:\:\:\rho}
-T^{\nu\mu}_{\:\:\:\:\rho}-T_{\rho}^{\:\:\:\:\mu\nu}\Big).
\end{equation}
Finally, it proves useful to define
\begin{equation}
 \label{Stensor}
S_\rho^{\:\:\:\mu\nu}=\frac{1}{2}\Big(K^{\mu\nu}_{\:\:\:\:\rho}+\delta^\mu_%
\rho \:T^{\alpha\nu}_{\:\:\:\:\alpha}-\delta^\nu_\rho\:
T^{\alpha\mu}_{\:\:\:\:\alpha}\Big).
\end{equation}
Note the antisymmetric relations $ T^{\lambda}{}_{\mu\nu} = -
T^{\lambda}{}_{\nu\mu}$ and $S_{\rho}{}^{\mu\nu} =
-S_{\rho}{}^{\nu\mu} $, as can be easily verified. Using these
quantities
 one can define the so called ``teleparallel Lagrangian'' as
\cite{Hayashi79,Maluf:1994ji,Arcos:2005ec}
\begin{equation}  \label{telelag}
L_T\equiv S_\rho^{\:\:\:\mu\nu}\:T^\rho_{\:\:\:\mu\nu}.
\end{equation}
In summary, in the present formalism, all the information
concerning the gravitational field is included in the torsion
tensor ${T}^\lambda_{\:\mu\nu}$, and the teleparallel Lagrangian
$L_T$ gives rise to the dynamical equations for the vierbein,
which imply the Einstein equations for the metric.

From the above discussion one can deduce that the teleparallel
Lagrangian arises from the torsion tensor, similar to the way the
curvature scalar arises from the curvature (Riemann) tensor. Thus,
one can simplify the notation by replacing the symbol $L_T$ by the
symbol $T$, which is the torsion scalar \cite{Linder:2010py}.

While in teleparallel gravity the action is constructed by the
teleparallel Lagrangian $L_T = T$, the idea of $f(T)$ gravity is
to generalize $T$ to a function $T+f(T)$, which is similar in
spirit to the generalization of the Ricci scalar $R$ in the
Einstein-Hilbert action to a function $f(R)$. In particular, the
action in a universe governed by $f(T)$ gravity reads:
\begin{eqnarray}
\label{action}
 I = \frac{1}{16\pi G }\int d^4x e
\left[T+f(T)+L_m\right],
\end{eqnarray}
where $e = \textrm{det}(e_{\mu}^A) = \sqrt{-g}$ and $L_m$ stands
for the matter Lagrangian. We mention here that since the Ricci
scalar $R$ and the torsion scalar $T$ differ only by a total
derivative \cite{Weinberg:2008}, in the case where $f(T)$ is a
constant (which will play the role of a cosmological constant) the
action (\ref{action}) is equivalent to General Relativity with a
cosmological constant.

Finally, we mention that throughout this work we use the common
choice for the form of the vierbien, namely
 \be \label{weproudlyuse} e_{\mu}^A={\rm
diag}(1,a,a,a).
 \ee
 It can be easily found that the family of vierbiens related to
(\ref{weproudlyuse})  through global Lorentz transformations, lead
to the same equations of motion. Note however that, as it was
shown in \cite{10101041}, $f(T)$ gravity does not preserve local
Lorentz invariance. Thus, one should in principle study the
cosmological consequences of a more general vierbien ansatz.
Definitely the subject needs further investigation and it is left
for a future work.

\subsection{Background $f(T)$ cosmology}
\label{fTcosm}

Let us now present the background cosmological equations in a
universe governed by $f(T)$ gravity. Variation of the action
(\ref{action}) with respect to the vierbein gives the equations of
motion \begin{widetext}
\begin{eqnarray}\label{eom}
e^{-1}\partial_{\mu}(eS_{A}{}^{\mu\nu})[1+f'({T})]
-e_{A}^{\lambda}T^{\rho}{}_{\mu\lambda}S_{\rho}{}^{\nu\mu}[1+f'({T})]
+
S_{A}{}^{\mu\nu}\partial_{\mu}({T})f''({T})-\frac{1}{4}e_{A}^{\nu}[T+f({T})]
= 4\pi G e_{A}^{\rho}\overset {\mathbf{em}}T_{\rho}{}^{\nu},
\end{eqnarray}
\end{widetext} where a prime denotes the derivative with respect to
$T$ and the mixed indices are used as in $S_A{}^{\mu\nu} =
e_A^{\rho}S_{\rho}{}^{\mu\nu}$. Note that the tensor $\overset
{\mathbf{em}}T_{\rho}{}^{\nu}$ on the right-hand side is the usual
energy-momentum tensor, in which we have added an overset label in
order to avoid confusion with the torsion tensor.

If we assume the background to be a perfect fluid, then the energy
momentum tensor takes the form
\begin{eqnarray}
\overset{\mathbf{em}}T_{\mu\nu} = pg_{\mu\nu} - (\rho +
p)u_{\mu}u_{\nu},
\end{eqnarray}
where $u^{\mu}$ is the fluid four-velocity. Note that we are
following the conventions of \cite{Weinberg:2008}, but with an
opposite signature metric. Thus, one can see that the equations
(\ref{eom}) lead to the background (Friedmann) equations
\begin{eqnarray}\label{background1}
&&H^2 = \frac{ 8\pi G }{3}\rho_m
-\frac{f({T})}{6}-2f'({T})H^2\\\label{background2}
&&\dot{H}=-\frac{ 4\pi G (\rho_m+p_m)}{1+f'(T)-12H^2f''(T)}.
\end{eqnarray}
 In
these expressions we have introduced the Hubble parameter
$H\equiv\dot{a}/a$, where a dot denotes a derivative with respect
to coordinate time $t$. Moreover, $\rho_m$ and $p_m$ stand
respectively for the energy density and pressure of the matter
content of the universe, with equation-of-state parameter
$w_m=p_m/\rho_m$. Finally, we have employed the very useful
relation
\begin{eqnarray}
T=-6H^2, \label{TH2}
\end{eqnarray}
which straightforwardly arises from evaluation of (\ref{telelag})
for the unperturbed metric.

Observing the form of the first Friedmann equation
(\ref{background1}), and comparing to the standard form, we deduce
that the second and third terms on the right hand side constitute
effectively the dark energy sector, which in general presents a
dynamical behavior. In particular, one can define the dynamical
dark energy (DDE) density as \cite{Linder:2010py}:
\begin{eqnarray}
\rhophi\equiv\frac{3}{8\pi
G}\left[-\frac{f({T})}{6}-2f'({T})H^2\right], \label{rhoDDE}
\end{eqnarray}
while its equation-of-state parameter reads: \be\label{wfT}
 w
=-\frac{f/T-f'(T)+2Tf''(T)}{\[1+f'(T)+2Tf''(T)\]\[f/T-2f'(T)\]}.
\ee
 Thus, in principle, any dynamical dark energy scenario, with a
given $\rhophi$ or a given $w$, has its $f(T)$ equivalent, and the
corresponding reconstruction procedure will be described in
section \ref{Reconstr}. Lastly, note that General Relativity is
recovered by setting $f(T)$ to a constant (which will play the
role of a cosmological constant), as expected.

\subsection{Linear Perturbations in $f(T)$ gravity}
\label{vacuumperturbations}

We now recall the first order perturbations of $f(T)$ gravity. We
refer the reader to \cite{Dent:2010va} for full details of the
calculation,  summarizing only the relevant results in this
section. The perturbed metric in Newtonian gauge with signature
$(+---)$ is written as
\begin{eqnarray}
\label{pertmetric}
 ds^2 = (1 + 2\psi)dt^2
-a^2(1-2\phi)\delta_{ij}dx^idx^j,
\end{eqnarray}
where the scalars $\phi$ and $\psi$ are functions of $\bx$ and
$t$. Writing the vierbein as
\begin{eqnarray}
e_{\mu}^A = \bar{e}_{\mu}^A + t_{\mu}^A,
\end{eqnarray}
where $\bar{e}_{\mu}^A$ is the unperturbed vierbein, for the above
metric we obtain
\begin{eqnarray}
\label{pert1} &&\bar{e}_{\mu}^0 = \delta_{\mu}^0\,\,\,\,\,
\bar{e}_{\mu}^a = \delta_{\mu}^aa\,\,\,\,\,
\bar{e}^{\mu}_0 = \delta^{\mu}_0 \,\,\,\,\, \bar{e}^{\mu}_a = \frac{\delta^{\mu}_a}{a}\\
&&t_{\mu}^0 = \delta_{\mu}^0\psi \,\,\,\,\, t_{\mu}^a
=-\delta_{\mu}^a a\phi \,\,\,\,\, t^{\mu}_0 =
-\delta_{0}^{\mu}\psi \,\,\,\,\, t^{\mu}_a =
\frac{\delta^{\mu}_a}{a}\phi, \ \ \ \ \label{pert2}
\end{eqnarray}
with indicial notation as stated at the beginning of section
\ref{model}. Moreover, unless otherwise indicated, subscripts zero
and one will generally denote respectively zeroth and linear order
values of quantities. We have also incorporated the additional simplifying assumption
that the perturbations $t_{\mu}^A$ are diagonal.  With these perturbations the determinant
becomes
\begin{eqnarray}
e = \textrm{det}(e_{\mu}^A) = a^3(1+\psi - 3\phi).
\end{eqnarray}

The perturbations of
the energy-momentum tensor are expressed as \cite{Dent:2010va}
\begin{eqnarray}
\label{T00pert}
\delta \overset {\mathbf{em}}T_0{}^0 &=& -\delta\rho_m\\
\delta \overset {\mathbf{em}}T_0{}^a &=& a^{-2}(\rho_m + p_m)(-\partial_a \delta u)\\
\delta\overset {\mathbf{em}} T_a{}^0 &=& (\rho_m +
p_m)(\partial_a\delta u) \label{Ta0pert}
\\
\delta\overset {\mathbf{em}} T_a{}^b &=& \delta_{ab}\delta p.
\label{Tabpert}
\end{eqnarray}

Pressureless matter implies that $\phi = \psi$, a relation that
simplifies the calculations significantly. Additionally, it proves
convenient to transform to Fourier space, where any first order
quantity $y$ is expanded as
\begin{eqnarray}
 y(t,\bx)=\int \frac{d^3k}{(2\pi)^\frac{3}{2}}
~y_k(t)e^{i\bk\cdot\bx}. \label{phiexpansion}
\end{eqnarray}
In what follows, the k-subscripts in a quantity denotes its
Fourier transformation.

Finally, from the above one can result to the following evolution
equation for the metric perturbation  \cite{Dent:2010va}:
\begin{eqnarray}
\label{phiddk} \ddot{\phi}_k+\Gamma(f) \dot{\phi}_k+\omega^2(f)
\phi_k=0,
\end{eqnarray}
where $\omega^2$ and $\Gamma$ are respectively given by
\begin{widetext}
\begin{eqnarray}
\label{omega2}
\omega^2(f)=\frac{\frac{3H^2}{2}+\dot{H}-\frac{f}{4}+\dot{H}f'-\(36H^4-48H^2\dot{H}\)f''+144H^4\dot{H}f'''}{1+f'-12H^2f''}
\end{eqnarray}
\begin{eqnarray}
\Gamma(f)=\frac{4H\[1+f'-\(12H^2+9\dot{H}\)f''+36H^2\dot{H}f'''\]}{1+f'-12H^2f''},
\end{eqnarray}
\end{widetext}
with primes denoting derivatives with respect to $T$.

\subsection{Growth of perturbations}
\label{growthpert}

 In order to study the growth of perturbations in a matter-only universe
 ($p_m=\delta p_m=0$)  we define as usual the overdensity $\delta$
as
 \be
 \label{deltadef}
 \delta\equiv\frac{\delta\rho_m}{\rho_m}.\ee
Thus, the relativistic version of the Poisson equation in $f(T)$
gravity reads \cite{Dent:2010va}
\begin{align}
\label{poisson}
&3H\(1+f'-12H^2f''\)\dot{\phi}_k\nonumber\\
&+\[\(3H^2+k^2/a^2\)\(1+f'\)-36H^4f''\]\phi_k\nonumber\\
&+4\pi G \rho\,\delta_k=0.
\end{align}
Equations (\ref{phiddk}) and (\ref{poisson}) can be used to evolve
the matter overdensity for a given $f(T)$ model.

\section{Dynamical dark energy and its $f(T)$
equivalent}\label{Reconstr}

In this work we are interested in constructing the $f(T)$ scenario
that effectively leads to the same background behavior with a
given cosmological model. We stress that this procedure has a full
generality, that is for any  dynamical dark energy (DDE) scenario
we can construct its $f(T)$ equivalent, as long as we know the
evolution of the dark energy density $\rhophi$. However, despite
the coincidence of the background behavior, the perturbations can
be used to distinguish the given DDE model from the corresponding
$f(T)$ reconstruction.

\subsection{Reconstructing the corresponding $f(T)$ family for any given Dynamical Dark Energy}
\label{ftqequiv}

In order to perform the aforementioned reconstruction at the
background level we start from the $\rhophi$ effective definition
(\ref{rhoDDE}). Using the fact that $\dot{f}=-12f'H\dot{H}$ (since
$T=-6H^2$), we can extract a differential equation for $f$ in
terms of $\rho_{\rm DDE}$, namely:
 \be
 \label{freconstruction}
 \dot{f}=\frac{\dot{H}}{H}\left(f+16\pi G
\rhophi\right) \ee
 The solution to this first-order differential
equation can be written in closed form as \be
 \label{fsol} f(H)=16\pi
G  H\int \frac{\rhophi }{H^2}dH+CH, \ee
 where $C$ is an
integration constant, and the corresponding $f(T)$ can be
straightforwardly obtained substituting $H$ by $\sqrt{-T/6}$. In
summary, relation (\ref{fsol}) describes a one-parameter family of
solutions, characterized by the parameter $C$, which by
construction  mimics perfectly the background evolution of the
given dynamical dark energy density. A self-consistency test for
this is that if one use the above $f(T)$ and the given $H$ to
calculate the equation-of-state parameter for the $f(T)$
equivalent through (\ref{wfT}), he finds exactly the same result
as using the conservation equation
$\dot{\rho}_{DDE}=-3H\rho_{DDE}(1+w)$ to calculate the
equation-of-state parameter for the given DDE model. Finally, for
the special case of General Relativity with a cosmological
constant ($\Lambda$) (where $\rhophi=\rho_\Lambda\equiv
\Lambda/(8\pi G)$), with $C=0$, the corresponding $f(T)$ model
reduces to the expected $f=-2\Lambda$.

Let us now make a comment on the parameter $C$. Interestingly, and
as expected, the $C$ term in (\ref{fsol}), which corresponds to a
term proportional to $(-T)^{1/2}$ in $f(T)$, does not have any
effect on the background dynamics, since it disappears from both
Friedmann equations (\ref{background1}) and (\ref{background2}),
and consequently from any other background-level quantity such as
luminosity distances or the equation of state. In other words, the
various specific scenarios of the family of solutions  given by
(\ref{fsol}), are indistinguishable from each other at the
background level.

\subsection{Using perturbations to distinguish between Dynamical Dark Energy and the corresponding f(T) family}
\label{pertgen}

In the previous subsection, we showed that the DDE and the
corresponding $f(T)$ models are perfectly indistinguishable at the
background level. However, this is not anymore the case if one
takes into account the perturbations, as we show in this
subsection. In particular, the perturbations can break the above
degeneracy, that is the growth history is different for the DDE as
well as for each member of the corresponding reconstructed $f(T)$
family.

Let us now proceed to the investigation of the perturbations. For
convenience, in the following we express the dimensional constant
$C$ of relation (\ref{fsol}) in terms of the dimensionless
quantity
\begin{eqnarray}
C_M\equiv -\frac{CH_i}{16\pi G\rho_{{\rm DDE},i}},
\end{eqnarray}
  where $H_i$ is the Hubble
parameter at an initial redshift $z_i$, that is $H_i\equiv
H(z=z_i)$, where we use the redshift $z$ as the independent
variable instead of the scale factor ($1 + z =a_0/a$ with $a_0$
the present scale-factor value). $\rho_{{\rm DDE},i}$ is the
initial energy density of the DDE.  If we assume $\rho_{{\rm
DDE},i}$ to be slowly varying during matter domination, then the
value of the integral in (\ref{fsol}) at the initial time
($z=z_i$) can be approximated by $-16\pi G \rho_{{\rm DDE},i}$,
and then $C_M$ can be physically interpreted from the fact that
the initial value of $f(H)$ at $z=z_i$ becomes $-16\pi G
(C_M+1)\rho_{{\rm DDE},i}$.

Amongst the individual members of a reconstructed family of $f(T)$
models, corresponding to a given DDE scenario, the growth
histories can be different for different choices of $C_M$. In
particular, the deviations in the growth history for different
values of $C_M$ are larger for smaller scales, while they diminish
for large scales. This behavior can be qualitatively understood
from equation (\ref{poisson}). Using  $\delta_f$ to denote the
matter perturbations under the reconstructed $f(T)$ scenarios, we
perform  in (\ref{poisson}) the transformation
 $f(T):f(T)\rightarrow F(T)+CH$, (where $F(T)$ is clearly
  the $C_M=0$ member of the family).
  We find that (\ref{poisson}) transforms to
\begin{align}
\label{poissonf}
&3H\(1+F'-12H^2F''\)\dot{\phi}_k\nonumber\\
&+\[\(3H^2+k^2/a^2\)\(1+F'\)-36H^4F''\]\phi_k\nonumber\\
&+4\pi G \rho_m\,\delta_{fk}=\frac{C}{12H}\frac{k^2}{a^2}\phi_k.
\end{align}

As we observe, the right-hand-side of the above equation indicates
the difference between the $\delta_f$ of the $C_M=0$ model and
that of the various $C_M\neq0$ ones. It is therefore clear that
for large scales ($k\rightarrow0$) the difference between the
$C_M\neq0$ and the $C_M=0$ models becomes small. In summary,
perturbations indeed uniquely distinguish a DDE scenario from the
different members of the corresponding reconstructed $f(T)$
family.

Finally, we mention that our results are similar in spirit to
\cite{HuSongSawicki} in the context of $f(R)$ gravity, where the
authors find that any expansion history of the universe can be
replicated by an one-parameter family of $f(R)$ models,
characterized by a parameter $B\propto f''(R)$, which provides a
variety of different behaviors for the different members of the
family. However, in the case of the present work there is a
difference, namely the various models of the family have a
relatively similar evolution even for large differences in the
parameter $C_M$.

\section{Application: $f(T)$ equivalent for quintessence}
\label{Quintappl}

In the previous section we described how we can construct $f(T)$
models that exhibit the same behavior at the background level with
any given dynamical dark energy (DDE) scenario. Furthermore, we
showed how the examination of perturbations can be used in order
to distinguish the DDE scenario from the various members of the
reconstructed $f(T)$ family, as well as these various members from
each other. In the present section, in order to provide specific
examples, we apply these procedures in the well-known dynamical
dark energy scenario of quintessence.

In the quintessence paradigm, the dark energy sector is attributed
to a homogeneous scalar field $\Phi$, while the matter sector is
described by a pressureless perfect fluid with energy density
$\rho_m$
\cite{RatraPeebles,ZlatevWangSteinhardt,SteinhardtWangZlatev:1999b,WangSteinhardt:1998}.
In the case of flat geometry the Friedman equation reads \be
\label{FR1}
 H^2=\frac{ 8\pi G
}{3}\left[\rho_m+\frac{\dot{\Phi}^2}{2}+V\(\Phi\)\right], \ee
where $V(\Phi)$ is the scalar-field potential.  In these models,
the field typically rolls down a very shallow potential,
eventually coming to rest when it can find a local minimum. In
particular, the evolution equation for $\Phi$ reads:
\begin{equation}
 \ddot{\Phi}+3H\dot{\Phi}+V'\(\Phi\)=0,
\end{equation}
while the corresponding one for $\rho_m$ is
\begin{equation}
\label{rhomevol}
 \dot{\rho}_m+3H\rho_m=0.
\end{equation}
Additionally, one can define the effective dark energy density and
pressure in field terms as
\begin{eqnarray}
\label{rhoDDEquint} && \rho_{\rm DDE}\equiv
\frac{\dot{\Phi}^2}{2}+V\(\Phi\)\\
&&p_{\rm DDE}\equiv \frac{\dot{\Phi}^2}{2}-V\(\Phi\),
\end{eqnarray}
and thus the corresponding dark-energy equation-of-state parameter
as
\begin{eqnarray}
\label{wDDEquint}
 w_{\rm DDE}\equiv \frac{p_{\rm DDE}}{ \rho_{\rm DDE}}=\frac{
\frac{\dot{\Phi}^2}{2}-V\(\Phi\)}{\frac{\dot{\Phi}^2}{2}+V\(\Phi\)},
\end{eqnarray}
which obviously present a dynamical behavior in general.
 Finally, in such a scenario, the growth of
perturbations is governed by the linearized Einstein equations
(see e.g. \cite{DentDutta}):
\begin{align}
&\dot{v}_f=-\phi_k\nonumber\\
&\dot{\delta}_k=3\dot{\phi}_k+\frac{k^2}{a^2}v_f\nonumber\\
&\ddot{\phi}_k+4H\dot{\phi}_k= 8\pi G \(\frac{\dot{\Phi}^2}{2}-V\)\phi_k+4\pi G  \(\dot{\Phi}\delta\dot{\Phi}-V'\delta\Phi\)\nonumber\\
&\delta\ddot{\Phi}=-3H\delta\dot{\Phi}-\(\frac{k^2}{a^2}+V''\)\delta\Phi+4\dot{\Phi}\dot{\phi}_k-2V'\phi_k,
\label{pertquintess}
\end{align} where $v_f$ is the velocity
potential of the matter fluid and primes denote derivatives with
respect to $\Phi$.

One last subject that has to be settled before proceeding forward
is the choice of the quintessence potential $V(\Phi)$. Amongst the
various ansatzes of the literature, in this work we focus on
three commonly studied cases, namely:
\begin{enumerate}
\item{The Pseudo-Nambu-Goldstone-Boson (PNGB) Model}: \be
\label{PNGB}
 V\(\Phi\)=A^4 \[1+\cos\(\Phi/f\)\],
  \ee where $A$ is
the energy scale of the potential, and $f$ is a symmetry
restoration scale. This model was first proposed in
\cite{Frieman}, while its cosmological applications were discussed in
\cite{kdutta}.

\item{The Exponential Model:}
 \be   V\(\Phi\)=A^4\,
\exp\(-\beta\Phi/M_{\rm Pl}\), \label{EXP}
 \ee
  where $M_{\rm Pl}$ is the Planck
mass.
 Exponential potentials arise in a variety of
contexts, such as higher dimensional gravitational theories,
moduli fields, and also non-perturbative effects such as gaugino
condensation (see e.g. \cite{Copeland:1997et} and references
therein).

\item{The Power Law Model:}
 \be
 \label{PowerLaw}
V\(\Phi\)=A^{4+n} \Phi^{-n}.
 \ee
  Power law potentials have been studied extensively in
\cite{RatraPeebles,ZlatevWangSteinhardt,Liddle:1998xm}, and have
been shown to arise in the context of SUSY in
\cite{Binetruy:1998rz}.
\end{enumerate}

The usual method of differentiating various types of DDE models is
to examine their phase-space behavior, focusing in particular on
the dark-energy equation-of-state parameter and its evolution
\cite{CaldwellLinder}. The various models can either fall into a
"thawing" or "freezing" category. The former is thusly named
because the field is only recently beginning to exhibit dynamical
behavior since it is being frozen at some distance from its
minimum due to Hubble friction, while the name of the latter
arises form the fact that it is dynamical throughout much of the
universe's history  but it becomes frozen into place at late times
(during the time of dark energy domination). Amongst the three
choices above, the first two are thawing models while the third one
is a freezing model.

\subsection{Reconstructing $f(T)$ for quintessence}
\label{fTquintess}

Let us now reconstruct an $f(T)$ cosmology that exhibits the same
behavior at the background level with a given quintessence
scenario. As described in subsection \ref{ftqequiv}, the
corresponding $f(T)$ family of models is given by relation
(\ref{fsol}), where in the quintessence case the dark energy
density $\rho_{\rm DDE}$ is given by (\ref{rhoDDEquint}), while
the cosmological equations close by  (\ref{FR1})-(\ref{rhomevol}).

In a non-trivial quintessence scenario, (\ref{fsol}) cannot be
solved analytically. Thus, in the following we perform a numerical
elaboration in the case of the three quintessence models described
above. In particular, knowing the evolution of $\rho_{\rm DDE}$
and of the effective dark-energy equation-of-state parameter $w$
for the three quintessence models, we numerically reconstruct the
corresponding $f(T)$ family using (\ref{fsol}), and then we
numerically extract the $w$-behavior using (\ref{wfT}).

For the three quintessence potentials we suitably choose the
parameter values in order for $w$ to lie within its observational
limits -1<$w_{\rm DDE}$<-0.9 for low redshifts \cite{Komatsu}. We
solve the cosmological equations determining the initial
conditions at a redshift $z_i$ deep inside the matter dominated
era, and we evolve the system until the density parameter of
matter $\Omega_{m}\(z\)\equiv 8\pi G \rho_m/(3H^2(z))$ becomes
equal to $\approx0.3$ at $z=0$, as it is required by observations.
For our numerical analysis, we work in units where $8\pi G=1$ and
$\rho_\Lambda=4/3$.
\begin{figure}[ht]
\begin{center}
            \epsfig{file=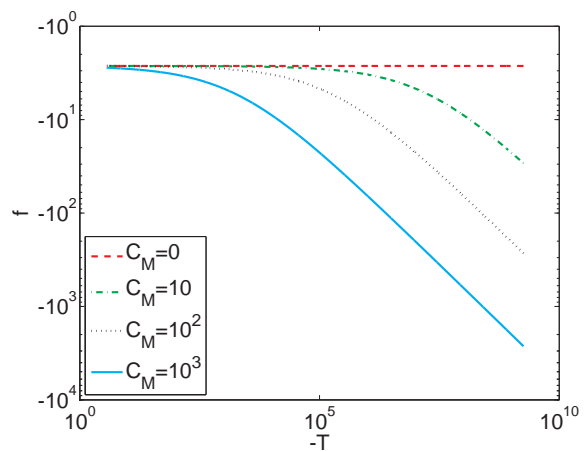,height=60mm}
    \caption
    {   \label{fPNGB} \textit{For members of the reconstructed f(T) family of models for the
    Pseudo-Nambu-Goldstone-Boson (PNGB) quintessence scenario
given by (\ref{PNGB}), characterized by four choices of the
parameter $C_M$. }}
    \end{center}
\end{figure}
\begin{figure}[ht]
\begin{center}
            \epsfig{file=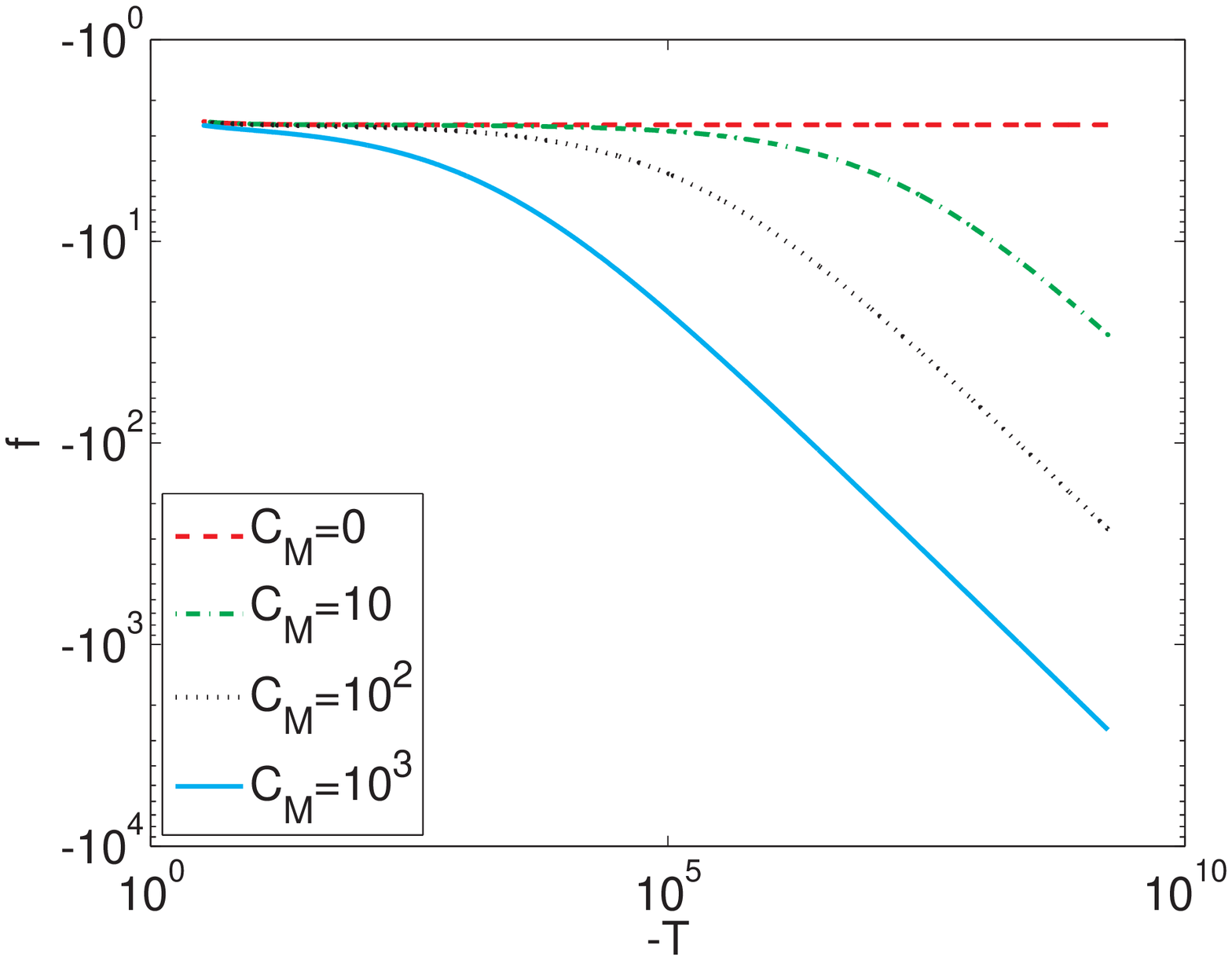,height=60mm}
    \caption
    {   \label{fEXP} \textit{
For members of the reconstructed f(T) family of models for the
exponential quintessence scenario given by (\ref{EXP}),
characterized by four choices of the parameter $C_M$.
  }}
    \end{center}
\end{figure}
\begin{figure}[ht]
\begin{center}
            \epsfig{file=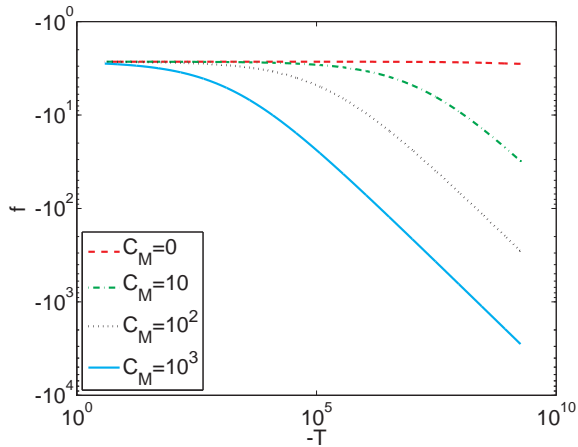,height=60mm}
    \caption
    {   \label{fPowerLaw} \textit{For members of the reconstructed f(T) family of models for the power-law
quintessence scenario given by (\ref{PowerLaw}), characterized by
four choices of the parameter $C_M$. }}
    \end{center}
\end{figure}

In Figures \ref{fPNGB}-\ref{fPowerLaw} we present the
reconstructed $f(T)$ for the three quintessence models, namely the
Pseudo-Nambu-Goldstone-Boson (\ref{PNGB}), the exponential
(\ref{EXP}), and the power-law (\ref{PowerLaw}) models,
respectively (such figures are easily created since we know
numerically $f(z)$ and $H(z)$, that is $T(z)$, and thus we obtain
$f(T)$). In each figure, we depict $f(T)$ for four members of the
$f(T)$ family of models, characterized by four choices of the
parameter $C_M$.
\begin{figure}[!]
\begin{center}
            \epsfig{file=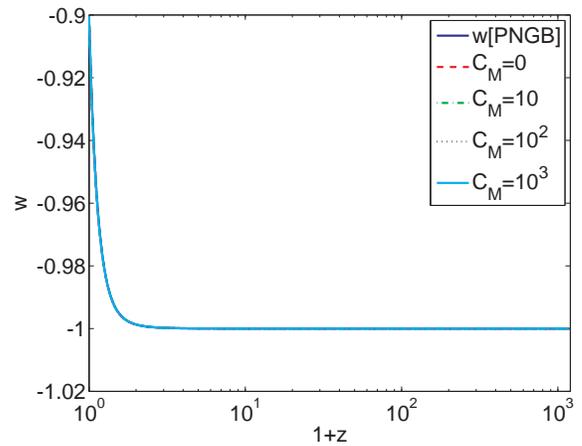,height=60mm}
    \caption
    {   \label{wPNGB} \textit{The evolution of the dark-energy
equation-of-state parameter $w(z)$, as a function of the redshift,
for the Pseudo-Nambu-Goldstone-Boson (PNGB) quintessence scenario
given by (\ref{PNGB}), as well as for four members of its
corresponding reconstructed $f(T)$ family of models (characterized
by four choices of the parameter $C_M$). }}
    \end{center}
\end{figure}
\begin{figure}[ht]
\begin{center}
            \epsfig{file=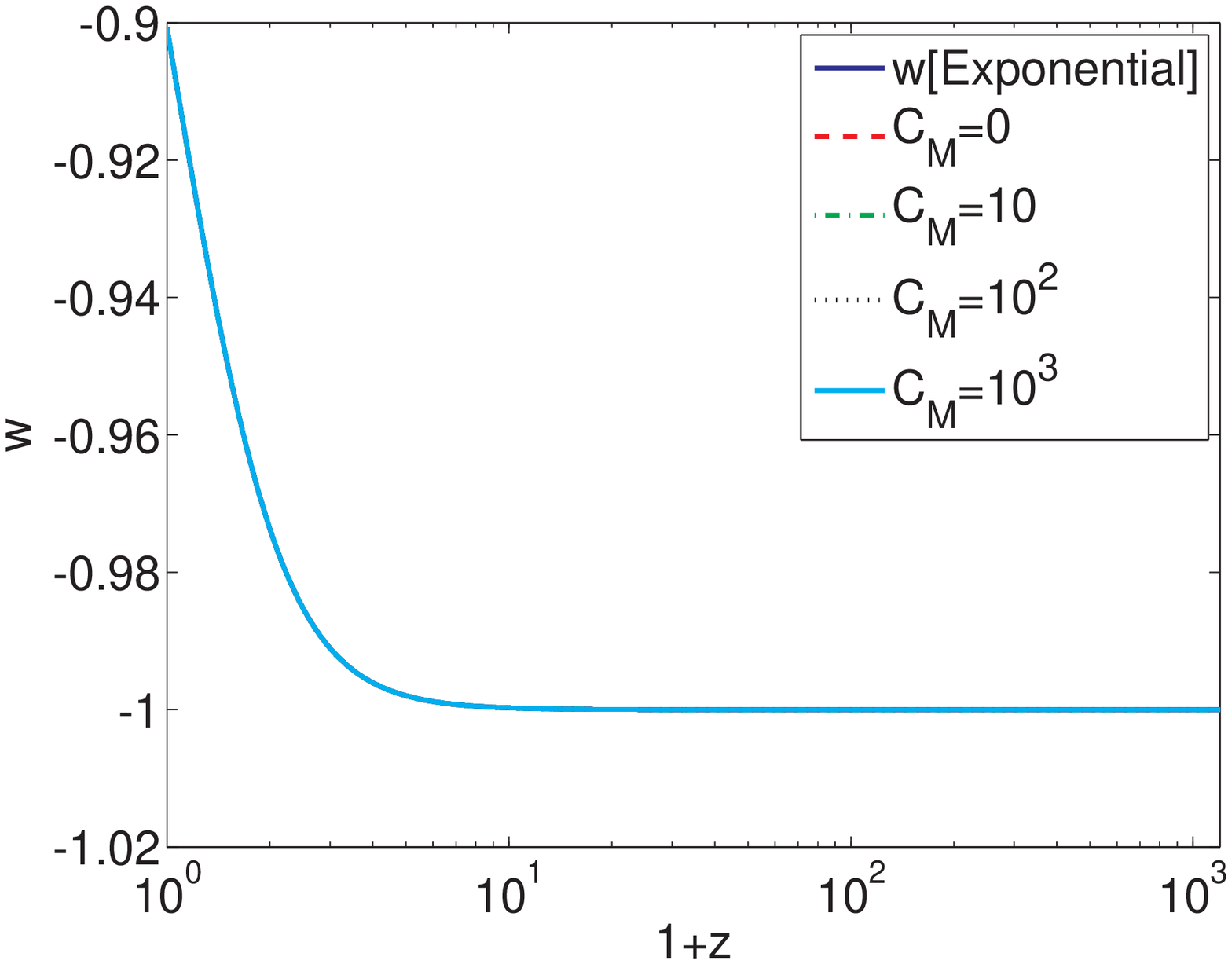,height=60mm}
    \caption
    {   \label{wEXP} \textit{
The evolution of the dark-energy equation-of-state parameter
$w(z)$, as a function of the redshift, for the exponential
quintessence scenario given by (\ref{EXP}), as well as for four
members of its corresponding reconstructed $f(T)$ family of models
(characterized by four choices of the parameter $C_M$). }}
    \end{center}
\end{figure}
\begin{figure}[ht]
\begin{center}
            \epsfig{file=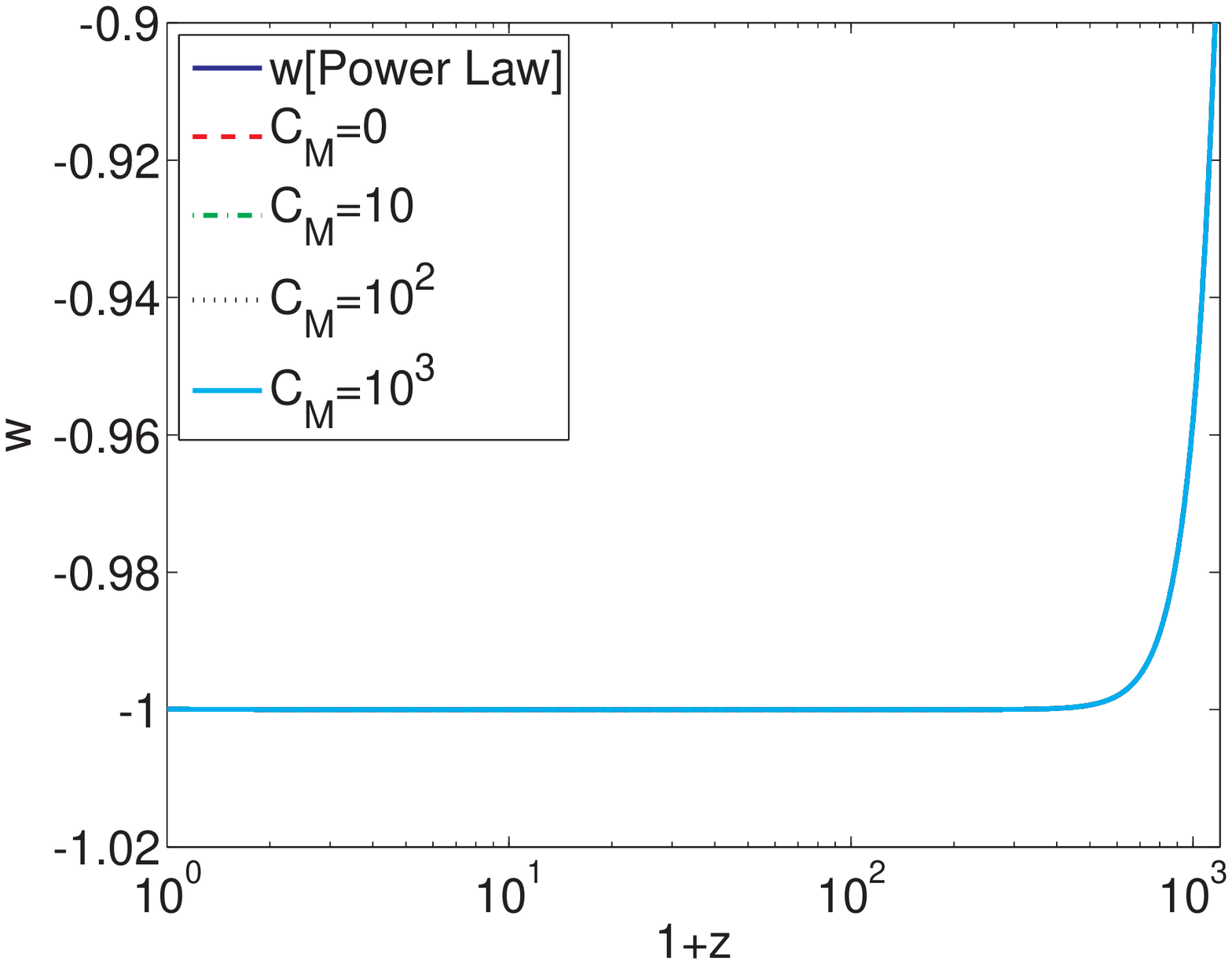,height=60mm}
    \caption
    {   \label{wPowerLaw} \textit{The evolution of the dark-energy equation-of-state parameter
$w(z)$, as a function of the redshift, for the power-law
quintessence scenario given by (\ref{PowerLaw}), as well as for
four members of its corresponding reconstructed $f(T)$ family of
models (characterized by four choices of the parameter $C_M$).}}
    \end{center}
\end{figure}

We mention that the $C$-term in (\ref{fsol}) behaves like a decaying
mode, as it falls of approximately as $a^{-3/2}$ during early
times, making the $C=0$ ($C_M=0$)  case an attractor for this
family of models. This is demonstrated in Figures
\ref{fPNGB}-\ref{fPowerLaw}, where it is clear that the $C_M\neq0$
models approach the $C_M=0$ one, for the reconstructions of all
three quintessence scenarios. Finally, note that we have focused
only on positive $C_M$-values, since negative values will produce
graphs which are symmetrically reflected about the
$f(z)=f(z=0)=-2\Lambda$ line.

Having reconstructed the $f(T)$ family of models that corresponds
to the three quintessence scenarios, in Figures
\ref{wPNGB}-\ref{wPowerLaw} we present the behavior of $w(z)$,
which is a basic observable, for each potential respectively,
according to (\ref{wDDEquint}). In each figure, we depict
additionally the $w(z)$ of the corresponding reconstructed $f(T)$
family of models, according to (\ref{wfT}), for four values of the
parameter $C_M$. As we observe, in all cases there is a perfect
overlap of the corresponding figures, which verifies the fact that
for background-level quantities, such is the dark-energy
equation-of-state parameter $w$, any dynamical dark energy
scenario is indistinguishable by construction from its
corresponding $f(T)$ equivalent family of models.

\subsection{Using perturbations to distinguish between quintessence and the corresponding f(T) family}

In the previous subsection, we applied our general
$f(T)$-reconstruction formalism of section \ref{ftqequiv} in the
case of quintessence dynamical dark energy scenario. In the
present subsection we investigate the perturbations following
subsection \ref{pertgen}, in order to distinguish between
quintessence and its equivalent family of models.

The perturbation evolution for quintessence is given by equations
(\ref{pertquintess}), while for the family of $f(T)$ models by
(\ref{poissonf}). We follow the numerical elaboration described in
the background case, and we additionally choose the initial
velocity of the metric perturbation to be at zero.
\begin{figure}[ht]
\begin{center}
            \epsfig{file=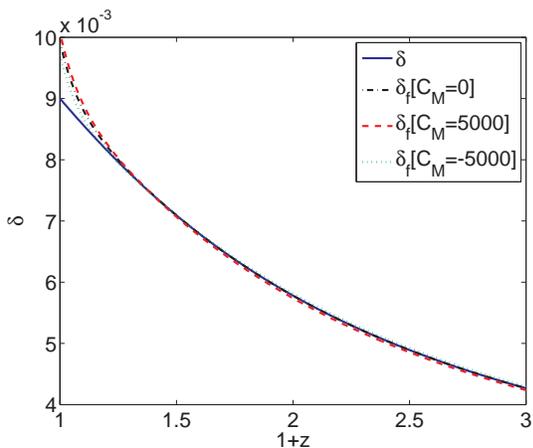,height=60mm}
    \caption
    {   \label{PNGB1000} \textit{The evolution of the matter overdensity
    $\delta$ as a function of the redshift $z$, on a scale of $k=10^{-3}h$ Mpc$^{-1}$,
    for the PNGB scenario given by (\ref{PNGB}),
as well as for three members of its corresponding reconstructed
$f(T)$ family of models (characterized by three choices of the
parameter $C_M$).}}
    \end{center}
\end{figure}

\begin{figure}[ht]
\begin{center}
    \epsfig{file=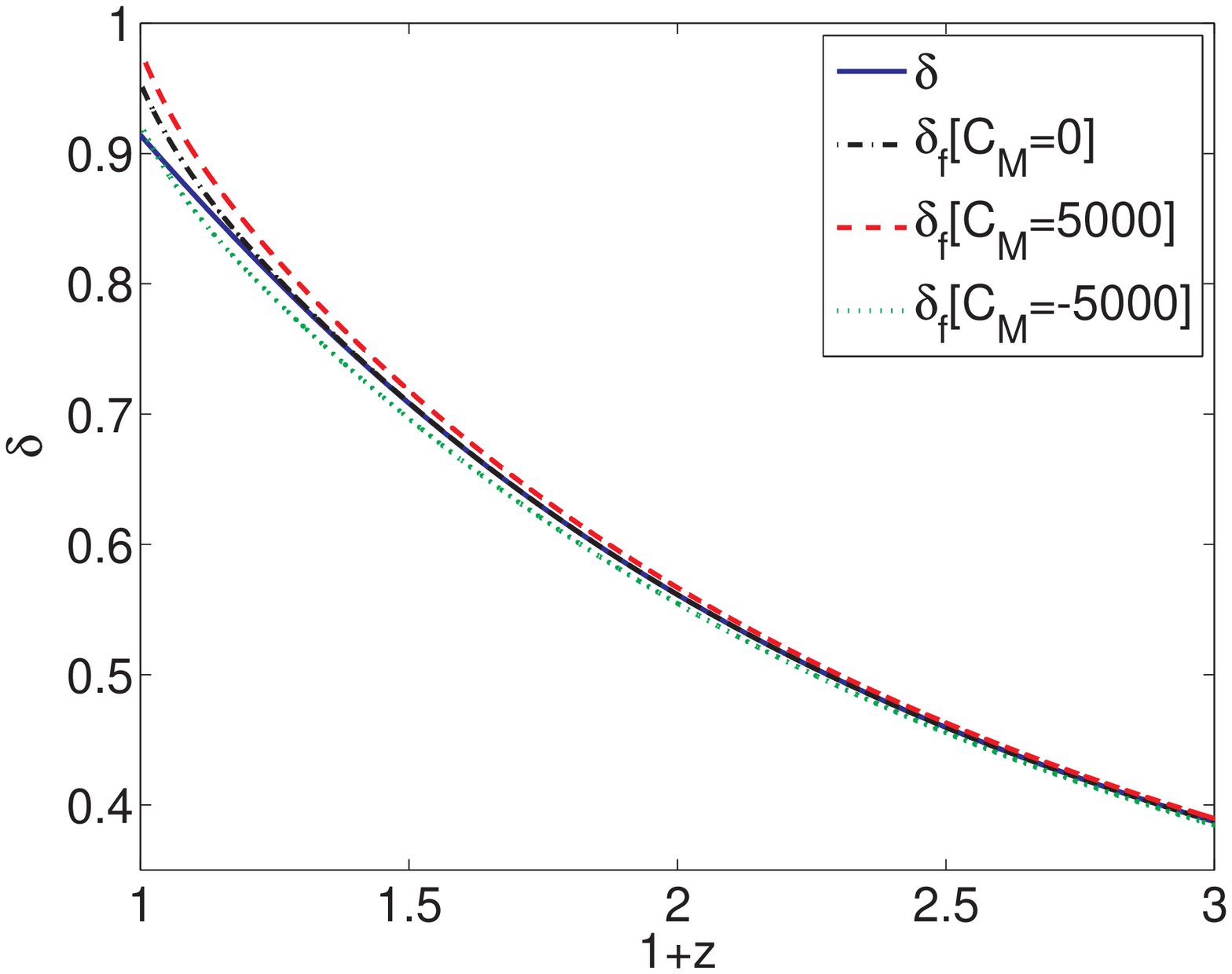,height=60mm}
    \caption
    {   \label{PNGB1} \textit{The evolution of the matter overdensity
    $\delta$ as a function of the redshift $z$, on a scale of $k=1h$ Mpc$^{-1}$,
     for the PNGB scenario given by (\ref{PNGB}),
as well as for three members of its corresponding reconstructed
$f(T)$ family of models (characterized by three choices of the
parameter $C_M$).  }}
 \end{center}
\end{figure}

\begin{figure}[ht]
\begin{center}
    \epsfig{file=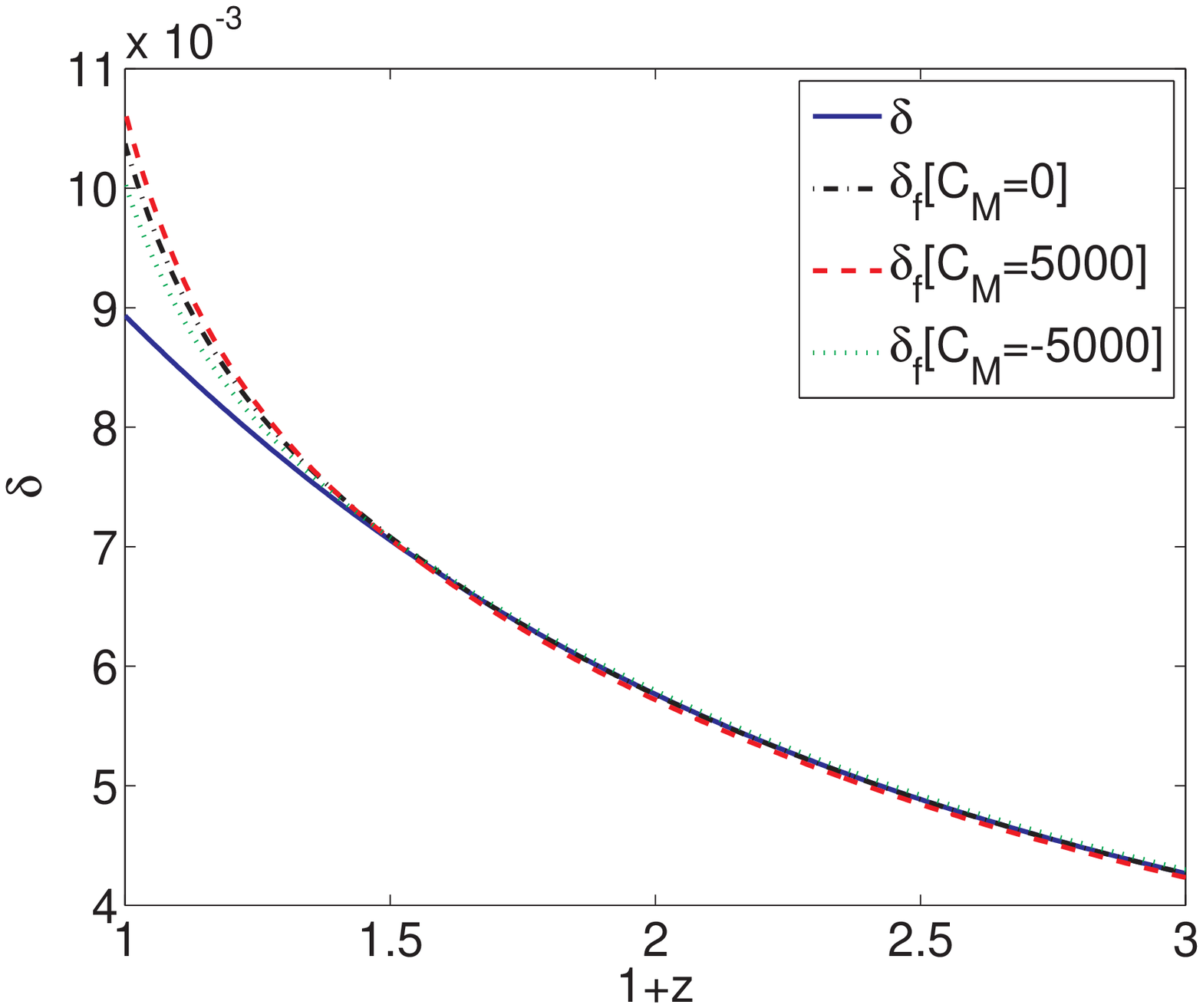,height=60mm}
    \caption
    {   \label{EXP1000} \textit{The evolution of the matter overdensity
    $\delta$ as a function of the redshift $z$, on a scale of $k=10^{-3}h$ Mpc$^{-1}$,
    for the exponential scenario given by (\ref{EXP}), as well as for four
members of its corresponding reconstructed $f(T)$ family of models
(characterized by four choices of the parameter $C_M$).  }}
  \end{center}
\end{figure}

\begin{figure}[ht]
\begin{center}
        \epsfig{file=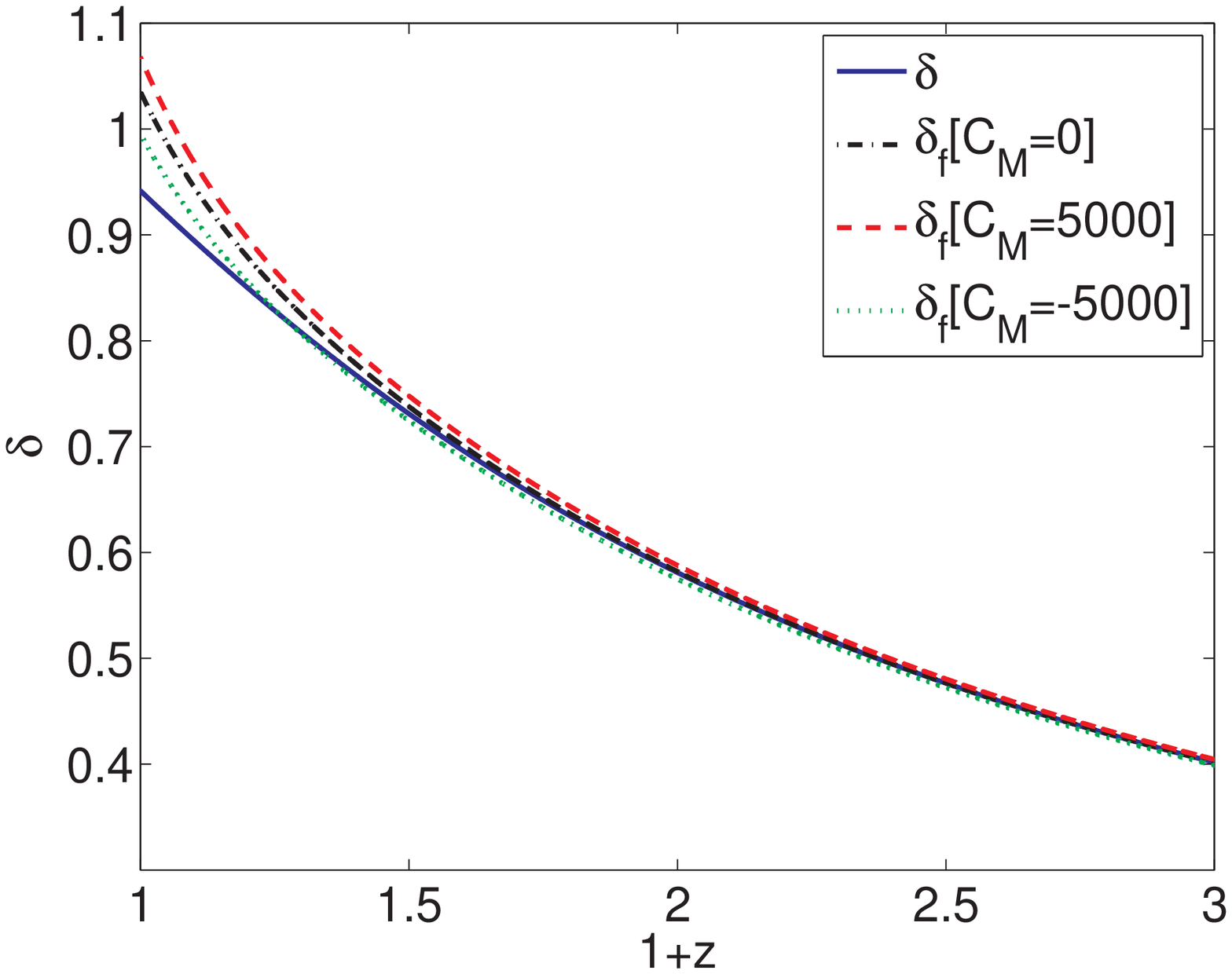,height=60mm}
    \caption
    {   \label{EXP1} \textit{The evolution of the matter overdensity
    $\delta$ as a function of the redshift $z$, on a scale of $k=h$ Mpc$^{-1}$,
   for the exponential scenario given by (\ref{EXP}), as well as for four
members of its corresponding reconstructed $f(T)$ family of models
(characterized by four choices of the parameter $C_M$). }}
 \end{center}
\end{figure}

\begin{figure}[ht]
\begin{center}
        \epsfig{file=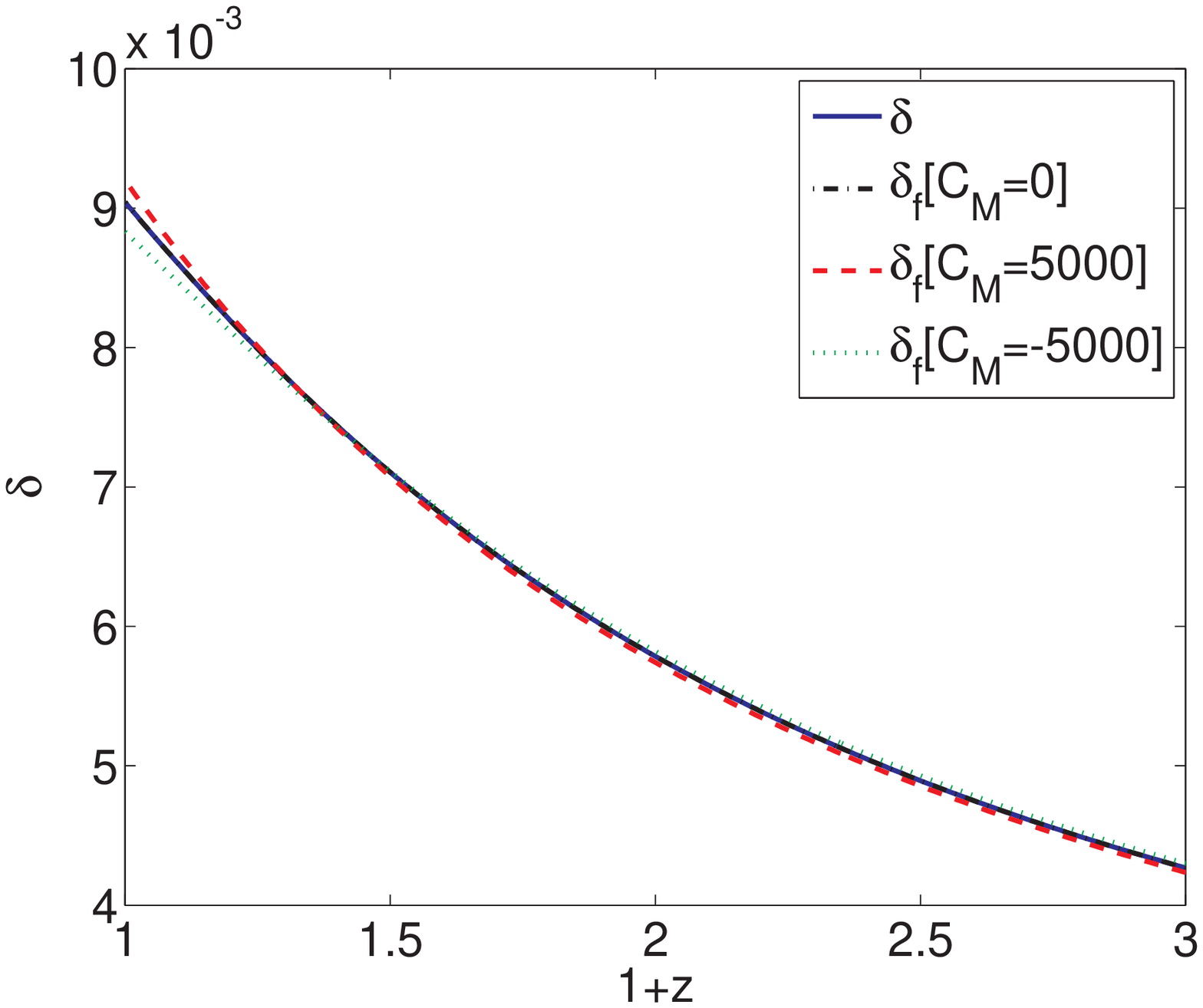,height=60mm}
    \caption
    {   \label{PowerLaw1000} \textit{The evolution of the matter overdensity
    $\delta$ as a function of the redshift $z$, on a scale of $k=10^{-3}h$ Mpc$^{-1}$,
    for the power-law scenario given by (\ref{PowerLaw}), as well as for three members of its corresponding reconstructed
$f(T)$ family of models (characterized by three choices of the
parameter $C_M$).   }}
\end{center}
\end{figure}

\begin{figure}[t]
\begin{center}
        \epsfig{file=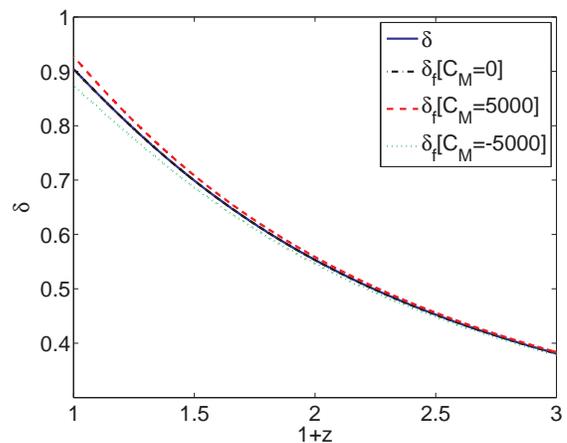,height=60mm}
    \caption
    {   \label{PowerLaw1} \textit{The evolution of the matter overdensity
    $\delta$ as a function of the redshift $z$, on a scale of $k=h$ Mpc$^{-1}$,
    for the power-law scenario given by (\ref{PowerLaw}), as well as for three members of its corresponding reconstructed
$f(T)$ family of models (characterized by three choices of the
parameter $C_M$).}}
 \end{center}
\end{figure}

In Fig.  \ref{PNGB1000} we demonstrate the evolution of the matter
overdensity $\delta$ as a function of the redshift, in the case of
Pseudo-Nambu-Goldstone-Boson (PNGB) quintessence scenario given by
(\ref{PNGB}), in a scale of $k=10^{-3}h$ Mpc$^{-1}$, while in Fig.
\ref{PNGB1} we provide the same graph but in a scale of $k=1h$
Mpc$^{-1}$. In the same Figures, we additionally depict the
$\delta$-evolution for three members of the corresponding
reconstructed $f(T)$ family, determined by three distinct values
of the parameter $C_M$ ($C_M=\pm5000$ and $C_M=0$). Similarly, in
Figures \ref{EXP1000} and \ref{EXP1} we provide the corresponding
plots for the exponential scenario given by (\ref{EXP}), while in
Figures \ref{PowerLaw1000} and \ref{PowerLaw1} we show the
corresponding graphs for the power-law scenario given by
(\ref{PowerLaw}).

From these Figures we observe that the evolution at the
perturbation level distinguishes the quintessence scenario from
its reconstructed equivalent family of $f(T)$ models, as well as
the various $f(T)$ family members from each other. Concerning the
scale, as discussed in subsection \ref{pertgen} the deviation in
$\delta$-evolution is strong for small scales and weak for large
scales. Concerning the redshift, for all the models the deviation
is negligible for large redshifts $z\gtrsim3$, which is expected
since the universe is matter dominated at that time and the
dark-energy sector (DDE or $f(T)$) plays no role in determining
the growth of perturbations. Moreover, note that that the shape of
the $\delta(z)$-curve differs fundamentally from that of the
$\delta_f(z;C_M)$-curves. Therefore, it is possible to choose a
value of $C_M$ such that $\delta$ and $\delta_f$ coincide at a
particular redshift, but they will not coincide at other redshifts
for this particular choice of $C_M$.

Finally, as we can see, in general the parameter $C_M$ has a very
weak impact on the splitting of $\delta$-evolutions, even for
small scales - the difference between the $C_M=5000$ and the
$C_M=-5000$ cases is modest in all the cases considered. This is
most likely a result of the attractor nature of the $C_M=0$
solution described in \ref{fTquintess}.

\section{Conclusions}
\label{conclusions}

In this work we investigated $f(T)$ cosmology in both the
background, as well as in the perturbation level, as a way to
mimic the behavior of a given model of dynamical dark energy.
$f(T)$ gravity is the extension of the ``teleparallel'' equivalent
of General Relativity, which uses the zero curvature
Weitzenb\"{o}ck connection instead of the  torsionless Levi-Civita
connection, in the same lines as $f(R)$ gravity is the extension
of standard General Relativity.

First, we presented the general formalism for reconstructing the
$f(T)$ scenario that effectively leads to the same behavior with
any given cosmological model, at the background level. As we
showed, for a given dynamical dark energy scenario, one can
reconstruct its equivalent, one-parameter family of models.
Although they exhibit completely indistinguishable background
behavior, the perturbations can break the above degeneracy.  That
is, the growth history is different for the given cosmological
scenario as well as for each member of the corresponding
reconstructed $f(T)$ family.

In order to present our results more transparently, we applied the
aforementioned general formalism in the well known case of
quintessence dynamical dark energy scenario. After reconstructing
the $f(T)$ family for three quintessence scenarios, we numerically
extracted the evolution of the dark-energy equation-of-state
parameter, which is a basic observable. Since it is a background
quantity it leads to identical behavior for quintessence as well
as for its $f(T)$ equivalent.

However, upon examining the perturbations, and in particular the
evolution of the matter overdensity, we do acquire a way to
distinguish between quintessence and its $f(T)$ equivalent family,
as well as between the infinite members of the reconstructed
$f(T)$ family. More specifically, the deviation is strong for
small scales and weak for large scales, and additionally it is
negligible for large redshifts ($z\gtrsim3$), since there the universe
is matter dominated with a negligible dark-energy sector.

In summary, $f(T)$ gravity can effectively mimic any dynamical
dark energy cosmological scenario at the background level. The
perturbation analysis can break this degeneracy leading to
rejection or acceptance of specific $f(T)$ models. These features
make $f(T)$ cosmology an interesting candidate for the description
of nature, which requires deeper examination.


\addcontentsline{toc}{section}{References}
\begin{thebibliography}{99}

\bibitem{union08}
  M.~Kowalski {\it et al.},
  Astrophys.\ J.\  {\bf 686}, 749 (2008).

\bibitem{perivol}
  L.~Perivolaropoulos and A.~Shafieloo,
  Phys.\ Rev.\  D {\bf 79}, 123502 (2009).

\bibitem{hicken}
  M.~Hicken {\it et al.},
  Astrophys.\ J.\  {\bf 700}, 1097 (2009).




\bibitem{Komatsu}
  E.~Komatsu {\it et al.},
  arXiv:1001.4538 [astro-ph.CO].



\bibitem{bao}
D.~J.~Eisenstein {\it et al.}  [SDSS Collaboration],
  Astrophys.\ J.\  {\bf 633}, 560 (2005).

\bibitem{percival}
W.~J.~Percival, S.~Cole, D.~J.~Eisenstein, R.~C.~Nichol, J.~A.~Peacock, A.~C.~Pope and A.~S.~Szalay,
  Mon.\ Not.\ Roy.\ Astron.\ Soc.\  {\bf 381}, 1053 (2007).



\bibitem{Samushia2007}
  L.~Samushia, G.~Chen and B.~Ratra,
  arXiv:0706.1963 [astro-ph].

\bibitem{Ettori}
  S.~Ettori {\it et al.},
  arXiv:0904.2740 [astro-ph.CO].


\bibitem{Wang}
  Y.~Wang,
  Phys.\ Rev.\  D {\bf 78}, 123532 (2008).

\bibitem{Samushia2009}
  L.~Samushia and B.~Ratra,
  Astrophys.\ J.\  {\bf 714}, 1347 (2010).


\bibitem{c7}
V. Sahni and A. Starobinsky, Int. J. Mod. Phy. D {\bf 9}, 373
(2000); P. J. Peebles and B. Ratra, Rev. Mod. Phys. {\bf 75}, 559
(2003).

\bibitem{varcc}
J.~Sola and H.~Stefancic,
Phys.\ Lett.\  B {\bf 624}, 147 (2005);
I.~L.~Shapiro and J.~Sola,
Phys.\ Lett.\  B {\bf 682}, 105 (2009).

\bibitem{RatraPeebles}
  B.~Ratra and P.~J.~E.~Peebles,
  Phys.\ Rev.\  D {\bf 37}, 3406 (1988).

\bibitem{ZlatevWangSteinhardt}
  I.~Zlatev, L.-M. ~Wang, and P.J. ~Steinhardt,
Phys.\ Rev.\ Lett.\ {\bf 82}, 896 (1999).

\bibitem{SteinhardtWangZlatev:1999b}
   P.J. ~Steinhardt, L.-M. ~Wang, and I.~Zlatev,
   Phys.\ Rev.\ D {\bf 59} 123504 (1999).

\bibitem{WangSteinhardt:1998}
  L.-M. ~Wang and P.J. ~Steinhardt,
  Astrophys. \ J. {\bf 508} 483 (1998).


\bibitem{quint}
C.~Wetterich, Nucl.\ Phys.\ B {\bf 302}, 668 (1988); A.~R.~Liddle
and R.~J.~Scherrer, Phys.\ Rev.\ D {\bf 59}, 023509 (1999)
  S.~Dutta and R.~J.~Scherrer,
  Phys.\ Rev.\  D {\bf 78}, 123525 (2008);
  Z.~K.~Guo, N.~Ohta and Y.~Z.~Zhang, Mod.\
Phys.\ Lett.\  A {\bf 22}, 883 (2007);
  S.~Dutta and R.~J.~Scherrer,
  Phys.\ Rev.\  D {\bf 78}, 083512 (2008);
  S.~Dutta and R.~J.~Scherrer,
  Phys.\ Lett.\  B {\bf 676}, 12 (2009);
  S.~Dutta, E.~N.~Saridakis and R.~J.~Scherrer,
Phys.\ Rev.\  D {\bf 79}, 103005 (2009);
  S.~Dutta, S.~D.~H.~Hsu, D.~Reeb and R.~J.~Scherrer,
  Phys.\ Rev.\  D {\bf 79}, 103504 (2009);
    T.~Chiba, S.~Dutta and R.~J.~Scherrer,
  Phys.\ Rev.\  D {\bf 80}, 043517 (2009);
  E.~N.~Saridakis and S.~V.~Sushkov,
  Phys.\ Rev.\  D {\bf 81}, 083510 (2010).


\bibitem{phant}
R. R. Caldwell, Phys.
Lett. B {\bf{545}}, 23 (2002); R.~R.~Caldwell, M.~Kamionkowski and
N.~N.~Weinberg, Phys. Rev. Lett. {\bf 91}, 071301 (2003); S.
Nojiri and S. D. Odintsov, Phys. Lett. B {\bf 562}, 147 (2003);
 Z.-K. Guo, Y.-S. Piao, and Y.-Z. Zhang, Phys. Lett. B {\bf 594},
247 (2004); E. Elizalde, S. Nojiri, and S.D Odintsov, \prd {\bf
70}, 043539 (2004); V. K. Onemli and R. P. Woodard, Phys.\ Rev.\ D
{\bf 70}, 107301 (2004); V. Faraoni, Class. Quant. Grav. {\bf 22},
3235 (2005); J. Kujat, R.J. Scherrer, and A.A. Sen, \prd {\bf 74},
083501 (2006);
 T. Chiba, \prd {\bf 73},
063501 (2006);
  E.~N.~Saridakis,
  Phys.\ Lett.\  B {\bf 676}, 7 (2009);
  M.~R.~Setare and E.~N.~Saridakis,
  JCAP {\bf 0903}, 002 (2009);
  E.~N.~Saridakis,
  Nucl.\ Phys.\  B {\bf 819}, 116 (2009).


\bibitem{quintom}
B.~Feng, X.~L.~Wang and X.~M.~Zhang, Phys.\ Lett.\  B {\bf 607},
35 (2005);
Z. K. Guo, {\it{et al.}}, Phys. Lett. B {\bf 608}, 177 (2005);
M.-Z Li, B. Feng, X.-M Zhang, JCAP, 0512, 002 (2005); B. Feng, M.
Li, Y.-S. Piao and X. Zhang, Phys. Lett. B {\bf 634}, 101 (2006);
 W. Zhao and Y.
Zhang, Phys. Rev. D {\bf73}, 123509 (2006);
  M.~R.~Setare and E.~N.~Saridakis,
  Phys.\ Lett.\  B {\bf 668}, 177 (2008);
  M.~R.~Setare and E.~N.~Saridakis,
  JCAP {\bf 0809}, 026 (2008);
    E.~N.~Saridakis and J.~M.~Weller,
  Phys.\ Rev.\  D {\bf 81}, 123523 (2010);
  Y.~F.~Cai, E.~N.~Saridakis, M.~R.~Setare and J.~Q.~Xia,
  Phys.\ Rept.\  {\bf 493}, 1 (2010).







\bibitem{coy}
T.~Chiba, T.~Okabe and M.~Yamaguchi,
  Phys.\ Rev.\  D {\bf 62}, 023511 (2000);
C.~Armendariz-Picon, V.~F.~Mukhanov and P.~J.~Steinhardt,
  Phys.\ Rev.\ Lett.\  {\bf 85}, 4438 (2000);
  C.~Armendariz-Picon, V.~F.~Mukhanov and P.~J.~Steinhardt,
  Phys.\ Rev.\ Lett.\  {\bf 85}, 4438 (2000);
T.~Chiba,
  Phys.\ Rev.\  D {\bf 66}, 063514 (2002).

\bibitem{unparticleDE}
D.~C.~Dai, S.~Dutta and D.~Stojkovic,
  Phys.\ Rev.\  D {\bf 80}, 063522 (2009).

\bibitem{Carroll:2003wy}
  S.~Capozziello, S.~Carloni and A.~Troisi,
  Recent Res.\ Dev.\ Astron.\ Astrophys.\  {\bf 1}, 625 (2003);
    S.~M.~Carroll, V.~Duvvuri, M.~Trodden and M.~S.~Turner,
  Phys.\ Rev.\  D {\bf 70}, 043528 (2004);
  S.~Nojiri and S.~D.~Odintsov,
   ``Modified gravity with negative and positive powers of the curvature:
  Phys.\ Rev.\  D {\bf 68}, 123512 (2003);
  A.~De Felice and S.~Tsujikawa,
  arXiv:1002.4928 [gr-qc].


\bibitem{GB}
  S.~Nojiri, S.~D.~Odintsov and M.~Sasaki,
  Phys.\ Rev.\  D {\bf 71}, 123509 (2005).

\bibitem{conformalgravDE}
  P.~D.~Mannheim,
  Prog.\ Part.\ Nucl.\ Phys.\  {\bf 56}, 340 (2006).



\bibitem{Nojiri:2005jg}
  S.~Nojiri and S.~D.~Odintsov,
  Phys.\ Lett.\  B {\bf 631}, 1 (2005);
  S.~Nojiri and S.~D.~Odintsov,
  J.\ Phys.\ Conf.\ Ser.\  {\bf 66}, 012005 (2007).

\bibitem{brane} P. Bin\'{e}truy, C. Deffayet, D. Langlois, Nucl. Phys. B {\bf565}, 269 (2000);
R.G. Cai, Y.G. Gong, B. Wang, JCAP {\bf0603}, 006 (2006); Y.G.
Gong, A. Wang, Class. Quantum Grav. {\bf23}, 3419 (2006);
  F.~K.~Diakonos and E.~N.~Saridakis,
  JCAP {\bf 0902}, 030 (2009).

  \bibitem{string}
  S.~Tsujikawa,
  arXiv:1004.1493 [astro-ph.CO].


\bibitem{holoext}
  S.~D.~H.~Hsu,
  Phys.\ Lett.\ B {\bf 594}, 13 (2004);
  M.~Li,
  Phys.\ Lett.\ B {\bf 603}, 1 (2004);
  E.~N.~Saridakis,
  Phys.\ Lett.\  B {\bf 660}, 138 (2008).

\bibitem{Horawa}
  P.~Horava,
  Phys.\ Rev.\  D {\bf 79}, 084008 (2009);
  G.~Calcagni,
  JHEP {\bf 0909}, 112 (2009);
    E.~Kiritsis and G.~Kofinas,
  Nucl.\ Phys.\  B {\bf 821}, 467 (2009);
  H.~Lu, J.~Mei and C.~N.~Pope,
Phys.\ Rev.\ Lett.\  {\bf 103}, 091301 (2009);
   E.~N.~Saridakis,
  Eur.\ Phys.\ J.\  C {\bf 67}, 229 (2010);
  X.~Gao, Y.~Wang, R.~Brandenberger and A.~Riotto,
Phys.\ Rev.\  D {\bf 81}, 083508 (2010); G.~Leon and
E.~N.~Saridakis,
  JCAP {\bf 0911}, 006 (2009);
  M.~i.~Park,
  JHEP {\bf 0909}, 123 (2009);
  S.~Dutta and E.~N.~Saridakis,
  JCAP {\bf 1001}, 013 (2010);
  C.~Germani, A.~Kehagias and K.~Sfetsos,
  JHEP {\bf 0909}, 060 (2009);
    E.~Kiritsis,
  Phys.\ Rev.\  D {\bf 81}, 044009 (2010);
  D.~Capasso and A.~P.~Polychronakos,
  JHEP {\bf 1002}, 068 (2010);
  S.~Dutta and E.~N.~Saridakis,
  JCAP {\bf 1005}, 013 (2010);
  G.~Koutsoumbas and P.~Pasipoularides,
  Phys.\ Rev.\  D {\bf 82}, 044046 (2010);
  M.~Eune and W.~Kim,
  arXiv:1007.1824 [hep-th];
  H.~B.~Kim and Y.~Kim,
  arXiv:1009.1201 [hep-th];
  H.~Kasari and T.~T.~Fujishiro,
  arXiv:1009.1703 [hep-th];
  T.~Harko, Z.~Kovacs and F.~S.~N.~Lobo,
  arXiv:1009.1958 [gr-qc];

\bibitem{Ferraro:2006jd}
  R.~Ferraro and F.~Fiorini,
  Phys.\ Rev.\  D {\bf 75}, 084031 (2007).

\bibitem{Bengochea:2008gz}
  G.~R.~Bengochea and R.~Ferraro,
  Phys.\ Rev.\  D {\bf 79}, 124019 (2009).

\bibitem{Linder:2010py}
  E.~V.~Linder,
  Phys.\ Rev.\  D {\bf 81}, 127301 (2010).

  \bibitem{ein28}
A. Einstein 1928, Sitz. Preuss. Akad. Wiss. p. 217; ibid p. 224;
  A.~Unzicker and T.~Case,
  {\it{Translation of Einstein's attempt of a unified field theory with
  teleparallelism}}
  arXiv:physics/0503046.

  \bibitem{Hayashi79}
K. Hayashi and T.  Shirafuji, Phys. Rev. D {\bf19}, 3524 (1979);
Addendum-ibid. {\bf24}, 3312 (1982).

\bibitem{Myrzakulov:2010vz}
  R.~Myrzakulov,
  arXiv:1006.1120 [gr-qc].

\bibitem{Myrzakulov:2010tc}
  R.~Myrzakulov,
  arXiv:1008.4486 [astro-ph.CO].

\bibitem{Yerzhanov:2010vu}
  K.~K.~Yerzhanov, S.~R.~Myrzakul, I.~I.~Kulnazarov and R.~Myrzakulov,
  arXiv:1006.3879 [gr-qc].


\bibitem{Yang:2010ji}
  R.~J.~Yang,
  arXiv:1010.1376 [gr-qc].

\bibitem{Wu:2010xk}
  P.~Wu and H.~Yu,
  arXiv:1007.2348 [astro-ph.CO].

\bibitem{Wu:2010mn}
  P.~Wu and H.~Yu,
  arXiv:1006.0674 [gr-qc].

\bibitem{Wu:2010av}
  P.~Wu and H.~W.~Yu,
  arXiv:1008.3669 [gr-qc];
  K.~Karami and A.~Abdolmaleki,
  arXiv:1009.2459 [gr-qc].


\bibitem{Bamba:2010iw}
  K.~Bamba, C.~Q.~Geng and C.~C.~Lee,
  arXiv:1008.4036 [astro-ph.CO].

\bibitem{Dent:2010va}
  J.~B.~Dent, S.~Dutta and E.~N.~Saridakis,
  arXiv:1008.1250 [astro-ph.CO].



\bibitem{Weitzenb23} Weitzenb\"{o}ck R., \emph{Invarianten Theorie}, (Nordhoff,
Groningen, 1923).

\bibitem{Maluf:1994ji}
  J.~W.~Maluf,
  J.\ Math.\ Phys.\  {\bf 35} (1994) 335.


\bibitem{Arcos:2005ec}
  H.~I.~Arcos and J.~G.~Pereira,
  Int.\ J.\ Mod.\ Phys.\  D {\bf 13}, 2193 (2004).

\bibitem{Weinberg:2008}
S. Weinberg, \emph{Cosmology}, Oxford University Press Inc., New
York, 2008.

 \bibitem{10101041}
  B.~Li, T.P.~Sotiriou, and J.D.~Barrow,
  arXiv:1010.1041 [gr-qc]




 \bibitem{HuSongSawicki}
  Y.~S.~Song, W.~Hu and I.~Sawicki,
  Phys.\ Rev.\  D {\bf 75}, 044004 (2007).


\bibitem{DentDutta}
  S.~Dutta and I.~Maor,
  Phys.\ Rev.\  D {\bf 75}, 063507 (2007);
J.~B.~Dent, S.~Dutta and T.~J.~Weiler,
  Phys.\ Rev.\  D {\bf 79}, 023502 (2009);
  J.~B.~Dent and S.~Dutta,
  Phys.\ Rev.\  D {\bf 79}, 063516 (2009).



\bibitem{Frieman}
J.A. Frieman, C.T. Hill, A. Stebbins, and I. Waga,
\prl {\bf 75}, 2077 (1995).

\bibitem{kdutta}
K. Dutta and L. Sorbo, \prd {\bf 75}, 063514 (2007);
 A. Abrahamse,
A. Albrecht, M. Barnard, and B. Bozek, \prd {\bf 77}, 103503
(2008);   R.~de Putter and E.~V.~Linder,
  JCAP {\bf 0810}, 042 (2008).


\bibitem{Copeland:1997et}
  E.~J.~Copeland, A.~R.~Liddle and D.~Wands,
  Phys.\ Rev.\  D {\bf 57}, 4686 (1998);
  A.~Albrecht and C.~Skordis,
  Phys.\ Rev.\ Lett.\  {\bf 84}, 2076 (2000);
  T.~Barreiro, E.~J.~Copeland and N.~J.~Nunes,
  Phys.\ Rev.\  D {\bf 61}, 127301 (2000).


\bibitem{Liddle:1998xm}
  A.~R.~Liddle and R.~J.~Scherrer,
  Phys.\ Rev.\  D {\bf 59}, 023509 (1999);
  E.~N.~Saridakis,
  Nucl.\ Phys.\  B {\bf 830}, 374 (2010).

\bibitem{Binetruy:1998rz}
  P.~Binetruy,
  Phys.\ Rev.\  D {\bf 60}, 063502 (1999);
  A.~Masiero, M.~Pietroni and F.~Rosati,
  Phys.\ Rev.\  D {\bf 61}, 023504 (2000).

\bibitem{CaldwellLinder}
 R.~Caldwell and E.V. ~Linder,
 Phys.\ Rev.\ Lett.\ {\bf{95}} 141301 (2005)





\end{thebibliography}
\end{document}